# Developing and Deploying Advanced Algorithms to Novel Supercomputing Hardware

Robert J. Brunner[1,2], Volodymyr V. Kindratenko[2], and Adam D. Myers[1]
1) Department of Astronomy, 2) National Center for Supercomputing Applications, University of Illinois at Urbana-Champaign
rb@astro.uiuc.edu, kindr@ncsa.uiuc.edu, adm@astro.uiuc.edu

*Abstract*-The objective of our research is to demonstrate the practical usage and orders of magnitude speedup of real-world applications by using alternative technologies to support high performance computing. Currently, the main barrier to the widespread adoption of this technology is the lack of development tools and case studies that typically impede non-specialists that might otherwise develop applications that could leverage these technologies. By partnering with the Innovative Systems Laboratory at the National Center for Supercomputing, we have obtained access to several novel technologies, including several Field-Programmable Gate Array (FPGA) systems, NVidia Graphics Processing Units (GPUs), and the STI Cell BE platform. Our goal is to not only demonstrate the capabilities of these systems, but to also serve as guides for others to follow in our path.

To date, we have explored the efficacy of the SRC-6 MAP-C and MAP-E and SGI RASC Athena and RC100 reconfigurable computing platforms in supporting a two-point correlation function which is used in a number of different scientific domains. In a brute force test, the FPGA based single-processor system has achieved an almost two orders of magnitude speedup over a single-processor CPU system. We are now developing implementations of this algorithm on other platforms, including one using a GPU. Given the considerable efforts of the cosmology community in optimizing these classes of algorithms, we are currently working to implement an optimized version of the basic family of correlation functions by using tree-based data structures. Finally, we are also exploring other algorithms, such as instance-based classifiers, power spectrum estimators, and higher-order correlation functions that are also commonly used in a wide range of scientific disciplines.

## I. Introduction

Application scientists, and the users of general-purpose computers have always desired greater processing performance, and these desires have historically been met with commodity-based computational systems. These systems, which have followed Moore's Law, an empirical relationship stating that processing power doubles every eighteen months, are no longer keeping pace with demand, however, as the requirements for constructing, cooling, and powering new central processing units (CPUs) have created an environment where Moore's Law no longer applies. As a result, the demands for increased performance must be met in a new manner.

Many commercial vendors have responded to this new reality by placing multiple cores on a single CPU, which require that applications be parallelized, via threading, or some other technique such as MPI or openMP, in order to fully utilize the available computational systems. These types of programming are often beyond the skills of most developers, however, resulting in under-utilized resources. An alternative approach is the application of accelerator technologies, such as reconfigurable computing or graphics programmable units (GPUs), that can be used to dramatically improve the performance of specific classes of algorithms.

In this paper, we present the results of our initial research on the application of accelerator technologies to basic algorithms, which, while drawn from the field of Cosmology, are applicable in many scientific and engineering domains. Here we focus on the specific task of implementing the two-point correlation function on two reconfigurable computing platforms: the SRC Computers SRC-6, and the SGI RASC RC100. First, we present the details of the two point correlation function, followed by the details of the two computational systems used in this current analysis. We then discuss the implementation details on both systems, and conclude with a discussion of our work and our planned future directions.

## II. BASIC ALGORITHM

Correlation functions are used extensively within the astronomy community to characterize the clustering of extragalactic objects. The two-point angular correlation function (TPACF) encodes the frequency distribution of separations between coordinate positions in a parameter space, as compared to randomly distributed coordinate positions across the same space. In astronomy applications, a common coordinate choice is the angular separations, θ, on the celestial sphere, which can be used to measure the angular two-point correlation function, which we will denote here as w(θ). Qualitatively, a positive value of w(θ) indicates that objects are found more frequently at angular separations of θ than would be expected for a randomly distributed set of coordinate points (i.e., a *correlation*). Similarly, w(θ)=*0* codifies a random distribution of objects, and w(θ) < *0* indicates an unexpected paucity of objects at separations of θ (i.e., an *anti-correlation*).

A preferred schema for estimating the two-point correlation function was derived by Landy and Szalay [1]. Estimating angular correlation functions generally requires computing histograms of angular separations between a particular set of positions in a data space. The positions in question might be the set of data points themselves, which we will denote *DD(θ)*, or a set of points that are randomly distributed in the same space as the data, which we will denote *RR(θ)*. Similarly the distribution of separations between the data sample and a set of random points, which we will denote *DR(θ)* can be calculated. Landy and Szalay's innovation was to suggest a way of minimizing the variance in estimates of w(θ) by estimating the correlation function as:

$$w(\theta) = \frac{DD(\theta) - 2DR(\theta) + RR(\theta)}{RR(\theta)}$$

Naively, calculation of the separation distributions (*DD, DR, RR*) for $N_D$ total points is an $O(N_D^2)$ problem, as it requires computing distances between all possible pairs of points in the data space. Additionally, as the variance of each of the separation distributions is inversely proportional to the number of points sampled, using a random sample that is $n_R$ times larger than the dataset is recommended. This guarantees that the finite size of the random sample introduces a contribution to the variance that is $n_R$ times smaller than the contribution from the data sample itself. To ensure the random points introduce fractional statistical imprecision compared to the natural limitations of the data, the random sample is usually constructed to contain $n_R$~100 times as many coordinate positions as the dataset. Computing the distribution of *all* separations for a random sample that is $n_R$ times larger than a dataset increases calculation complexity by a factor of $n_R^2$. As modern astronomical data sets can contain many millions of positions, complexity can grow rapidly. One might therefore prefer to create $n_R$ unique random samples of comparable size to the dataset, and then average the separation distributions over these individual realizations, thus reducing the complexity introduced by sampling across the random realizations to $n_R$. Fortunately, statistical precision is not reduced by such an approach [1]. Equation 1 can then be written:

$$w(\theta) = \frac{n_R \cdot DD(\theta) - 2 \sum_{i=0}^{n_R-1} DR_i(\theta)}{\sum_{i=0}^{n_R-1} RR_i(\theta)} + 1$$

where $n_R$ is the number of sets of random points.

Astronomical measurements are usually made in a spherical coordinate system, with the coordinate positions expressed as Right Ascension and Declination (i.e., latitude and longitude) pairs. The separation, θ, between any two positions *p* and *q* in such a coordinate system can be determined by first converting the spherical coordinates to Cartesian coordinates, and computing θ via:

$$\theta = \arccos(p \times q) = \arccos(x_p x_q + y_p y_q + z_p z_q)$$

Observationally, determining w(θ) requires binning the separation distributions at some angular resolution Δθ. The binning schema implemented by astronomers is typically logarithmic, as cosmological clustering patterns approximate power-law behavior across a wide range of angular scales. Each decade of angle in the logarithmic space is divided equally between *k* bins, meaning that there are *k* equally-logarithmically-spaced bins between, for example, 0.01 and 0.1 arcminutes. The bin edges are then defined by $10^{j/k}$, where *j=-∞,…,-1,0,1,…+∞*, and the following formula can be used to find the integer bin number for angular separation θ:

$$\text{bin} = \text{integer}[k(\log_{10} \theta - \log_{10} \theta_{min})]$$

where $\theta_{min}$ is the smallest angular separation that can be measured.

Note that the binning schema described above requires the calculation of *arccos* and *log* functions,

which are computationally expensive. If only a small number of bins are required, a faster approach is to project the bin edges to the pre-arccosine "dot product" space and search in this space to locate the corresponding bin. Since the bin edges are ordered, an efficient binary search algorithm can be used to quickly locate the corresponding bin in just $\log_2 M$ steps, where $M$ is the total number of bins. The bin edges in the "dot product" space can be computed as follows:

$$\text{binedge}_j = \cos(10^{\log_{10}\theta_{min} + j/k}), j = 0, ..., M$$

Theoretically, each possible angular separation falls in a unique bin, and w(θ) can thus be uniquely determined for any distribution of points. Angular coordinates, however, as measured by modern astronomical surveys, are typically precise to ~0.1 arcseconds (e.g., [2]). The definitions of the bin edges are absolute; but the θ values have some built-in tolerance. Expressing θ values to different numbers of decimal places, therefore, can cause separations to drift between bins, affecting an estimate of w(θ), but *not rendering that estimate incorrect*. Any differences in the estimates of w(θ) that are introduced by imprecision in measured coordinates are, in most instances, completely undetectable, as variations in the random samples used to estimate *DR(*θ) and *RR(*θ) will usually dominate this numerical imprecision.

For reference, we made a C implementation of the TPACF algorithm suitable for execution on the conventional microprocessor platform. The computational core of the algorithm is a subroutine which calculates binned separation distributions for either DD(θ) (and RR(θ)) or DR(θ) style counts, depending on the input parameters. Henceforth, we will often refer to DD(θ) (and RR(θ)) or RR(θ) counts as "autocorrelations" and DR(θ) counts as "cross-correlations". To avoid confusion, we will always refer to the full angular two-point autocorrelation as the TPACF, or simply as w(θ). Initially, the data points are loaded/converted from spherical to Cartesian coordinates and the autocorrelation function for this dataset, DD(θ), is computed. Random points are then loaded/converted one realization at a time. For each random set, the autocorrelation for the random dataset, RR(θ) and the cross-correlation between the data points and the random set, DR(θ), are computed.

### III. RECONFIGURABLE COMPUTING

#### A. General concepts

Reconfigurable computing (RC) [3] based on the combination of conventional microprocessors and field-programmable gate arrays (FPGAs) has reached the point where select scientific kernels can be substantially accelerated with the ease of a C/FORTRAN style of programming. FPGAs, which were invented in the mid-80's by Ross Freeman [4], have been around for quite some time. However, until the late 1990's they did not achieve sufficient gate density or the functional capability to support non-trivial double-precision floating-point operations as required by many scientific kernels. And only recently have high-level languages (HLL) and code development tools become available that hide the complexity of the hardware design techniques involved in a typical FPGA design implementation cycle. As a result, high-performance reconfigurable computing (HPRC), or reconfigurable supercomputing, is a relatively recent technology that is actively growing as new algorithms and kernels are implemented on this new computational paradigm.

HPRC technology combines the advantages of the coarse-grain process-level parallel processing provided by conventional multiprocessor systems with the fine-grain instruction-level parallel processing available in FPGAs. The technological promise associated with this advantage has prompted traditional high-performance computing (HPC) vendors, such as SGI® and Cray Inc., to introduce several commercial HPRC products. In addition, newcomers to the HPC arena, such as SRC Computers, Inc. and Nallatech Ltd., have emerged with viable solutions. All of these systems consist of a traditional computer based on general-purpose processors and a separate "accelerator" component built around an FPGA. While similar in the basic concept, these individual solutions differ in the design of the accelerator component, the coupling between the accelerator component and the general-purpose computer system, and the access and control software.

In addition, several tools have recently been developed that can compile code written in a high-level language directly into the hardware circuitry description. For example, the SRC Computers Carte™ and the Mitrionics™ Mitrion SDK are two such development suites. Dividing the code between the general-purpose processor and the FPGA accelerator is not a trivial task [5], and is still the responsibility of the software developer. However, once the decision about the software/hardware code partitioning is made, the code developer can implement the hardware side of the chosen algorithm on the selected FPGA platform using the appropriate toolset, and the software side (to be executed on the general-purpose processor) using conventional code development techniques. To date, numerous codes have been successfully ported to various reconfigurable supercomputing platforms, including, to name just a few, molecular dynamics [6], various linear algebra solvers [7], and bioinformatics [8].

*B. SRC Computers SRC-6*

The SRC-6 MAPstation [9] used in this work consists of a standard dual 2.8 GHz Intel Xeon motherboard and a MAP Series E processor interconnected with a 1.4 GB/s low-latency 4-port Hi-Bar switch. The SNAP™ Series B interface board is used to connect the CPU board to the Hi-Bar switch. The SNAP plugs directly into a CPU board's DIMM memory slot.

The MAP Series C/E processor modules contain two user-controlled FPGAs, one control FPGA, and associated memory. There are six banks (A-F) of on-board memory (OBM); each bank is 64 bits wide and 4 MB deep, for a total of 24 MB. The programmer is explicitly responsible for application data transfer to and from these memory banks via SRC programming macros invoked from within the FPGA application. There is an additional 4 MB of dual-ported memory dedicated to the transfer of data between the two FPGAs. This memory can also be used as two additional OBM banks, named G and H.

The two user FPGAs in the MAP Series C are Xilinx Virtex-II XC2V6000 FPGAs. They each contain 67,584 4-input lookup tables, 67,584 flip flops, 144 dedicated 18x18 integer multipliers, and 324 KB of internal dual-ported block RAM (BRAM). The two user FPGAs in the MAP Series E are Xilinx Virtex-II Pro XC2VP100 FPGAs; they each contain 88,192 4-input lookup tables, 88,192 flip flops, 444 dedicated 18x18 integer multipliers, and 999 KB of internal dual-ported block RAM. The static FPGA clock rate of 100 MHz is enforced by the SRC programming environment.

*C. SGI RASC RC100*

In our experiments, we use a standalone, single-module SGI Altix 350 system [10] with a single dual-blade chassis containing one RC100 blade. The SGI Altix 350 is a dual-1.4 GHz Intel Itanium 2 system with 4 GBs of physical memory. An RC100 blade is attached to the host system via a NUMAlink 4 interconnect (see Fig. 1).

The RC100 is SGI's third-generation Reconfigurable Application-Specific Computing hardware module. It contains two computational FPGAs, two peer-to-peer I/O (TIO) ASICs and a special-purpose FPGA for loading bitstreams onto the computational FPGAs. The two user FPGAs are connected to the corresponding TIO ASICs via the Scaleable System Ports (SSPs). In addition, 10 QDR SRAM memory modules, each up to 8 MB, can be installed, in configuration up to five banks per FPGA chip. The SRAMs are configured as two banks to match the NUMAlink 4 channel bandwidth (3.2 Gbyte/sec) to the memories (2x1.6 Gbyte/sec).

The two user FPGAs are Xilinx Virtex 4 LX200 (XC4VLX200-FF1513-10) chips. Each chip contains 200,448 logic cells, 336 Block RAM/FIFOs with 6,048 kbits of total Block RAM, 96 DSP48 slices, and 960 user I/O pins. The maximum clock frequency of the chips, as implemented in the RC100, is 200 MHz. A portion of each chip is allocated to the RASC Core Services logic with the rest of the logic allocated to the user algorithm block.

*D. Software development challenge*

Software development for HPRC platforms consists of two separate paths: host microprocessor application development and embedded FPGA kernel implementation. The host microprocessor application is developed in C programming language with provisions made to accommodate a call to a compute kernel executed on an FPGA. Typically, a $3^{rd}$ party API is used to orchestrate the use of the FPGA. Such an API includes subroutines to allocate FPGA resource, transfer control to it during the application execution, send/receive data between the host and embedded processor systems, etc. API characteristics vary from system to system.

On the embedded processor side, code for the FPGA platform is written using a software development framework/tool set provided by either the hardware platform manufacturer (as in the case of SRC-6), or by a $3^{rd}$ party software vendor (as in the case of SGI RASC). An effective use of either of these tools requires a detailed knowledge of the underlying hardware architecture as well as proficiency with the vendor-specific programming language/style.

The Carte programming environment [11] for the SRC-6 MAPstation is highly integrated, as all of the compilation targets can be generated via a single makefile. The two possible compilation targets are a software-emulated "debug" version to verify functional correctness and the final version that contains the embedded circuit bitmap that runs on the actual hardware FPGA. The debug version is useful for code testing before the final time-intensive hardware place and route step. The Intel icc compiler is used to generate both the CPU-only debug executable and the CPU-side of the combined

CPU/MAP executable. The SRC MAP compiler is invoked by the makefile to produce the hardware description of the FPGA design for the final combined CPU/MAP target executable. This intermediate hardware description of the FPGA design is passed to the Xilinx Integrated Synthesis Environment (ISE) place and route tools, which produce the FPGA bit file. Finally, the linker is invoked to combine the CPU code and the FPGA hardware bit file(s) into a unified executable.

The Mitrion SDK [12] provides the framework in which we implemented our algorithm on the RC100 platform. Mitrion-C source can be verified for correctness using a functional simulator/debugger provided with the SDK. The Mitrion-C compiler will generate VHDL code from the Mitrion-C source and setup the instance hierarchy of the RASC FPGA design that includes the user algorithm implementation, the RASC Core Services, and configuration files necessary to implement the design. The design is then synthesized using the Xilinx suite of synthesis and implementation tools. In addition to the bitstream generated by the Xilinx ISE, two configuration files are created: one describes the algorithm's data layout and streaming capabilities to the RASC Abstraction Layer (bitsream configuration file) and the other describes various parameters used by the RASC Core Services. These files, together with the bitstream file, are required by the device manager to communicate with the algorithm that is implemented on the FPGA.

## IV. IMPLEMENTING COSMOLOGY CODES ON HPRC PLATFORMS

The dataset and random samples we use to calculate w(θ) are the sample of photometrically classified quasars and the random catalogs first analyzed by [13]. We use 100 random samples ($n_R$=100). The dataset, and each of the random realizations, contains 97178 points ($N_D$=97,178). We use a binning schema with five bins per decade (k=5), $θ_{min}$=0.01 arcminutes, and $θ_{max}$=10,000 arcminutes. Thus, angular separations are spread across 6 decades of scale and require 30 bins (M=30). Covering this range of scales requires the use of double-precision floating-point arithmetic as single-precision floating-point numbers do not have the precision sufficient to compute angular values smaller than 1 arcminute.

### A. Two-point angular correlation on SRC-6

The reference C implementation was written with the autocorrelation and cross-correlation functions coded as a single subroutine. On the other hand, it can be advantageous to implement these two functions separately when porting the implementation to the SRC-6 platform, because we have two different FPGA chips in the two MAPs, and each implementation can be targeted to best match the available chip resources to the function properties.

In our first implementation, the "autocorrelation" subroutine was written in MAP C targeting the MAP Series C reconfigurable processor module. The design occupies both FPGAs and makes use of all available OBM banks. The code implemented on the primary chip is responsible for transferring bin boundaries, bin values, and the sample to be processed into OBM banks. Bin boundaries and existing bin values are transferred first; they are mirrored by each FPGA to the on-chip memory. Sample points are transferred next, they are distributed across all 6 OBM banks and permissions are given to the secondary chip to access only one half of the memory banks. The workload is then split equally between the two FPGAs. Once the entire sample of coordinate points is processed and the results are obtained on both chips, they are merged on the primary chip and streamed out to the host microprocessor.

The computational core of the TPACF algorithm is implemented as a nested loop that closely follows the reference C implementation, with one important exception. We note that the MAP C compiler attempts to pipeline only the innermost loop of a code. The binary search loop, used to find the bin edges that a coordinate-point-separation lies between, is the innermost loop in the reference C implementation. Pipelining this loop alone does not lead to an efficient FPGA implementation, as multiple clock cycles would have to be spent to locate the bin that needs to be updated. Therefore, it is more advantageous to fully unroll this loop and let the MAP C compiler pipeline the next innermost loop instead. Thus, instead of running a binary search loop, we implement the analog of a series of multiple if-else statements by using a macro provided by SRC. This way, a new result can be computed upon each iteration of the pipelined loop, thus achieving a substantial improvement in efficiency of the overall computation. Moreover, there is enough space on each chip to execute two such calculations simultaneously. Thus, the overall execution time of this design is proportional to $N^2/8$ where N is number of points in the sample being processed: the overall algorithm complexity is ~$N^2/2$ and the execution is divided between two chips with 2 simultaneous calculations per chip.

The "cross-correlation" subroutine was written in MAP C targeting the MAP Series E reconfigurable processor module. As with the autocorrelation subroutine, the code implemented on the primary chip is responsible for transferring the bin boundaries, the bin values, and the sample to be processed into OBM banks. As before, the workload is then split equally between two FPGAs, and the results are assembled at conclusion. MAP Series E FPGAs are larger and they allow one to implement 3 simultaneous distance calculation/bin mapping cores per chip. Thus, overall execution time of this design is proportional to $N^2/6$: the overall algorithm complexity is $N^2$ and the execution is divided between two chips with 3 simultaneous calculations per chip.

A closer look at the numerical precision of the bin boundaries used in the calculations shows that just 12 digits after the decimal point (41 bits of the mantissa) are sufficient to provide the required precision used in this particular application. Thus, instead of comparing full double-precision floating-point numbers, it is sufficient to compare only the first 12 digits after the decimal point. Significant FPGA resource savings can be achieved when using this custom-designed data type: over 27% of SLICEs per single bin mapping core of 31 comparison operators. As a result, in our second implementation the autocorrelation subroutine was extended from two simultaneous distance calculation/bin mapping cores per chip to four such cores; the overall execution time of this design is proportional to $N^2/16$. Also, the cross-correlation subroutine was extended from three to five computational cores per chip; the overall execution time of this design is proportional to $N^2/10$.

Close examination of the reference C implementation reveals that there is a task-level parallelism in the algorithm that we have not yet exploited. In particular, once a random sample is loaded from the disk, computations of the autocorrelation and cross-correlation functions involving this dataset are entirely independent and may thus be executed simultaneously. Moreover, while calculations are executed on the MAPs using one dataset, the next random data dataset can be loaded and converted by the microprocessor. We can easily modify our reference C code to take advantage of running three simultaneous execution threads via OpenMP. One such thread is responsible for reading in a sample, the second thread is responsible for executing the autocorrelation subroutine, and the third thread is responsible for executing the cross-correlation subroutine. Thus, in our third implementation [14] we achieved coarse-grain task-level parallelism using multithreaded execution on the conventional microprocessor platform in addition to the fine-grain instruction-level parallelism implemented via the direct hardware execution of the core algorithms on two MAPs.

Such a simplistic code penalization technique however leads to a load-unbalanced implementation since the execution time of the autocorrelation subroutine is shorter than the execution time of the cross-correlation subroutine. As a result, one of the MAP processors is idle about 18% of the time. A better approach is to divide the datasets to be processed into smaller segments and distribute the workload between the two MAP processors dynamically [15]. We implemented the job scheduler as a simple loop that iterates over all the segments/segment pairs (tasks) to be processed and schedules each such task for the execution on the first freed MAP processor. Each dataset is divided into 5 equally sized segments, although higher granularity is desirable for larger datasets. As a result, each MAP processor is fully utilized and a better overall speedup of the code executed on the dual-MAP system, as compared to a single CPU system, is achieved. In our final implementation, the dual-MAP TPACF algorithm outperforms the SRC-6 host system (a single 2.8 GHz Intel Xeon microprocessor) by a factor of **96x** [15].

### B. Two-point angular correlation on RASC RC100

The autocorrelation/cross-correlation subroutine was re-written in the Mitrion-C language, targeting the RC100 platform [16]. Structurally, the Mitrion-C implementation of the computational core resembles the reference C implementation. On each iteration of the outer loop, a new point is loaded from the off-chip memory and is used throughout the entire inner loop execution. On each iteration of the inner loop, a new point is loaded from the off-chip memory and is used in the computation of the dot product. Once the dot product is computed, the bin to which it belongs is identified and updated. Actual bin boundaries are hardcoded in this initial implementation; they are saved as a vector that is stored on the chip. This storage mechanism allows the Mitrion-C compiler to fully unroll the bin array search 'for' loop into a 32-stage deep pipeline. Once the bin index is found, the corresponding bin value is incremented by one. Initially, bin values are stored as a vector and set to zero. As with the bin boundaries, this storage mechanism allows the Mitrion-C compiler to fully unroll the bin update 'foreach' loop into a wide pipeline. Since the bin search and bin update loops can be fully unrolled, the compiler is able to produce a fully pipelined inner loop implementation, thus generating an efficient overall algorithm implementation in which a new result is produced on each clock cycle. After all the calculations are performed, the resulting bin values are written back to the off-chip memory. From there, they are copied to the host memory via a RASC library call.

On the RC100 platform, the Mitrion-C processor has access to just two off-chip memory banks; each such bank is 128-bit wide and a few Megabytes deep. This memory is single-ported as far as the memory read access from the user application is concerned. Since the point coordinates are stored as double-precision floating-point numbers, each point requires 3x64 bits of storage space. In order to avoid a pipeline stall while reading each data point, we distribute coordinate points between the two memory banks. This data storage schema allows simultaneous access to the coordinate values of a single data point within a single clock cycle. Since the data structure used in the reference C implementation is not compatible with the required memory usage model, the data on the host system has to be reformatted before being sent to the RC100 module memory. This adds to the total algorithm execution time, but the overhead is minimal.

Note that even though our Mitrion-C implementation of the computational kernel is setup as a cross-correlation subroutine, we also use this kernel to compute the autocorrelation function. The only drawback of using this subroutine to compute the autocorrelation is that the final bin counts need to be divided by two and the overall execution time is twice the time actually required to compute the autocorrelation. This approach was necessary since, at the time this work was done, Mitrion-C did not provide an efficient way to implement variable length loops as required for the inner loop in a true autocorrelation implementation.

In the first implementation, 47% of slices and 50% of hardware multipliers were occupied by the final design. The Mitrion-C compiler uses 16 hardware multipliers to implement a single double-precision floating-point multiplier. This requires 48 (50% of all available) hardware multipliers to implement a 3D dot product on the Xilinx Virtex-4 VLX200 chip, leaving 48 hardware multipliers unused. As a result, there were sufficient FPGA resources left on the chip to implement an additional compute engine per chip. The modifications to the Mitrion-C source code necessary to implement the two compute engines per chip were trivial: two points are loaded instead of one from the off-chip memory on each iteration of the outer loop and two separate dot product/bin mapping/bin update paths were instantiated inside the inner loop. The results were merged at the end before being stored in the off-chip memory. No modifications to the previously used data storage or subroutine call from the host processor were required.

As with the SRC-6 dual-MAP implementation, since there are, similarly, two FPGAs in the RC100 blade, the cross- and autocorrelation kernels were executed simultaneously, one subroutine per FPGA. In this case, however, we do not need to deal with load balancing issue since both FPGAs execute the cross-correlation subroutine only, thus resulting in the same execution time on each chip. When comparing the execution time of this final implementation with the execution time of the same code on the Altix 350 host system (a single 1.4 GHz Intel Itanium 2), overall speedup of **25x** was achieved [16].

## V. CONCLUSIONS AND FUTURE WORK

In this paper, we have presented our results in porting the two-point angular correlation function to two reconfigurable computing platforms: the SRC Computers SRC-6, and the SGI RASC RC100. In both cases, we generated an optimal implementation of the standard brute-force algorithm that produced substantial performance improvements when compared to a standard CPU-based system. These implementations, however, required a substantial decomposition of the basic algorithm, highlighting the importance of our research as guides to the development other algorithms to this new class of computational systems.

With the success of our initial efforts, we are now extending our work to utilize additional accelerator platforms and other classes of algorithms. While we will continue to explore the efficacy of the next-generation reconfigurable platforms, such as the SRC Computers SRC-7, we are also exploring new technologies, such as the NVidia G80 graphics programmable unit. Similarly, while the two-point angular correlation function has produced remarkable insight into the use of accelerator technology, we are now decomposing other algorithms, such as a power spectrum measurement and specific class of machine learning algorithms, such as instance-based classifiers, in order to implement them on alternative computational platforms and expand our knowledge of optimal techniques for programming accelerator technologies.

ACKNOWLEDGMENT

This work was funded by NASA grant NNG06GH15G. We would like to thank David Caliga, Dan Poznanovic, and Jeff Hammes, all from SRC Computers Inc., for their help and support with SRC-6 system. We also would like to thank Tony Galieti, Matthias Fouquet-Lapar, and Dick Riegner, all from SGI, for their help and support with the SGI system and Stefan Möhl and Jace Mogill, both from Mitrionics AB, for their help with Mitrion SDK.